\documentclass[10pt,onecolumn,conference]{IEEEtran}
\usepackage{graphicx}
\usepackage{amsmath}
\usepackage{amsfonts}
\usepackage{float}
\usepackage{epstopdf}
\usepackage{bm}
\usepackage{array}
\usepackage{multirow}
\usepackage{multicol}
\usepackage{mathtools}
\usepackage{caption}
\usepackage{subcaption}
\usepackage{dblfloatfix}
\usepackage{tabularx}
\usepackage{multirow}
\usepackage{lipsum}
\usepackage{fancyhdr}
\usepackage{colortbl}
\usepackage[table]{xcolor}

\graphicspath{ {\Images} } 

\AtBeginDocument{%
  \providecommand\BibTeX{{%
    \normalfont B\kern-0.5em{\scshape i\kern-0.25em b}\kern-0.8em\TeX}}}

\begin{document}

\title{Experimental Study of Outdoor UAV Localization and Tracking using Passive RF Sensing}
\author{Udita Bhattacherjee, Ender Ozturk, Ozgur Ozdemir, Ismail Guvenc, Mihail L. Sichitiu, Huaiyu Dai\\
\IEEEauthorblockA{Department of Electrical and Computer Engineering, North Carolina State University, Raleigh, NC
}\\
Email: {\tt \{ubhatta,eozturk2,oozdemi,iguvenc,mlsichit,hdai\}@ncsu.edu}}






\maketitle

\begin{abstract}
  Extensive use of unmanned aerial vehicles (UAVs) is expected to raise privacy and security concerns among individuals and communities. In this context, detection and localization of UAVs will be critical for maintaining safe and secure airspace in the future. In this work, Keysight N6854A radio frequency (RF) sensors are used to detect and locate a UAV by passively monitoring the signals emitted from the UAV. First, the Keysight sensor detects the UAV by comparing the received RF signature with various other UAVs' RF signatures in the Keysight database using an envelope detection algorithm. Afterward, time difference of arrival (TDoA) based localization is performed by a central controller using the sensor data, and the drone is localized with some error. To mitigate the localization error, implementation of an extended Kalman filter~(EKF) is proposed in this study. The performance of the proposed approach is evaluated on a realistic experimental dataset. EKF requires basic assumptions on the type of motion throughout the trajectory, i.e., the movement of the object is assumed to fit some motion model~(MM) such as constant velocity (CV), constant acceleration (CA), and constant turn (CT). In the experiments, an arbitrary trajectory is followed, therefore it is not feasible to fit the whole trajectory into a single MM. Consequently, the trajectory is segmented into sub-parts and a different MM is assumed in each segment while building the EKF model. Simulation results demonstrate an improvement in error statistics when EKF is used if the MM assumption aligns with the real motion.
\end{abstract}



\begin{figure*}[b!]
     \centering
     \hspace{-1.3cm}
     \begin{subfigure}[t]{0.35\textwidth}
         \centerline{
         \includegraphics[trim= 0cm -1.5cm 0cm 0cm, clip,scale=0.325]{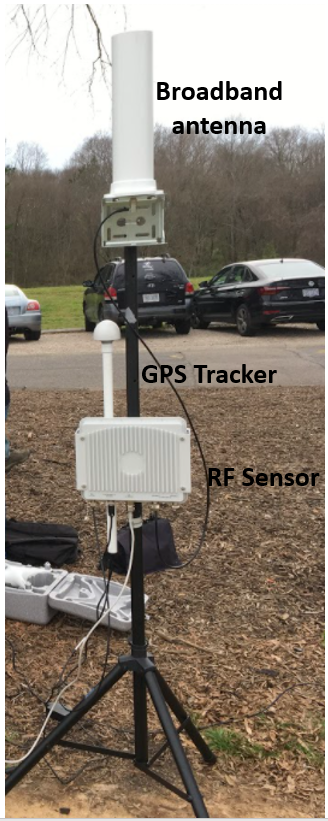}}
         \caption{ }
         \label{fig:RF_sensor}
     \end{subfigure}
     \hspace{-1cm}
     \begin{subfigure}[t]{0.35\textwidth}
         \centerline{
         \includegraphics[trim= 0cm -1.2cm 0cm 0cm, clip,scale=0.4]{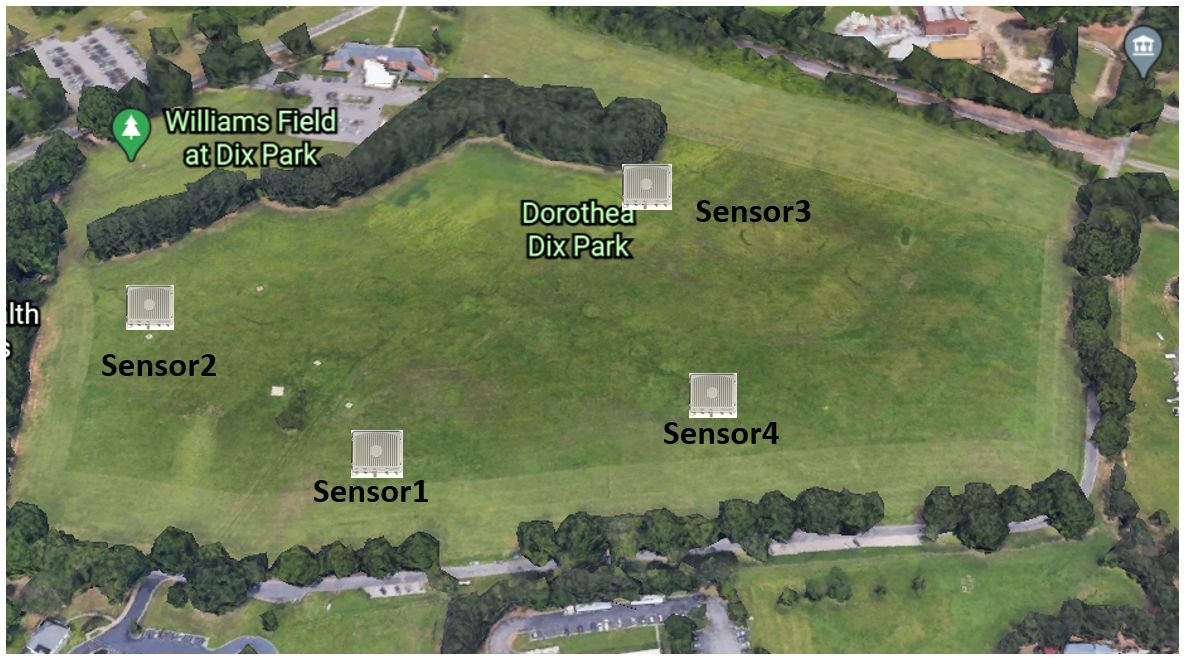}}
         \caption{ }
         \label{fig:Google_Earth}
     \end{subfigure}
     \hspace{.8cm}
    \begin{subfigure}[t]{0.35\textwidth}
         \centerline{\includegraphics[scale=0.49]{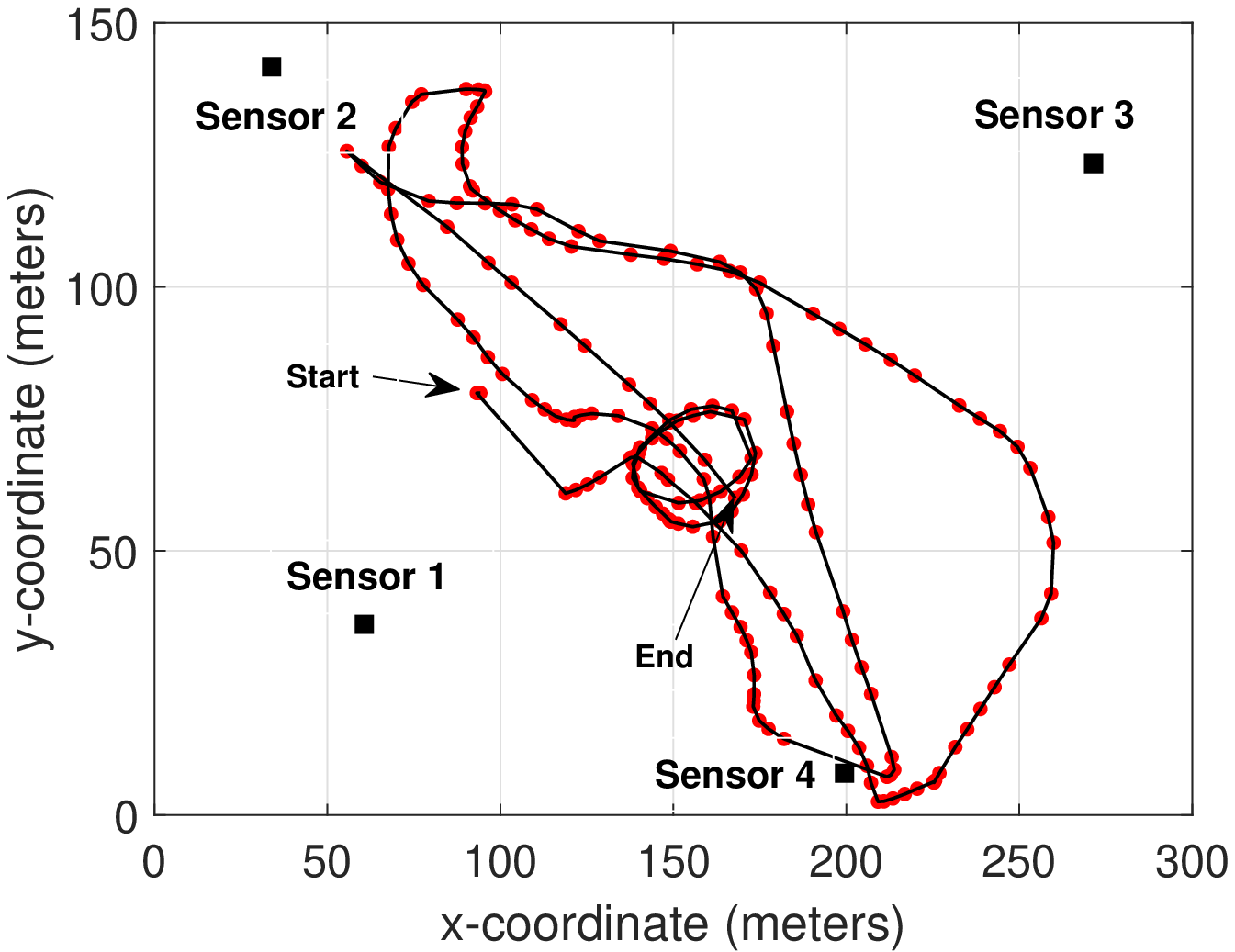}}
         \caption{ }
         \label{fig:UAV_Trajectory}
     \end{subfigure}
        \caption{(a) Keysight RF sensor
N6841A. b) Google Earth snapshot of Dorothea Dix park and location of the sensors. (c) Experimental UAV trajectory relative to the sensor locations.}
        \label{fig:Google_Eath_RF_Sensor}
\end{figure*}

\section{Introduction}
Unmanned aerial vehicles (UAVs), commonly known as drones, are expected to be an important component in next-generation wireless  systems~\cite{menouar2017uav}. UAVs have multiple commercial, government, and personal uses, ranging from goods delivery, live video surveillance, photography, among others. On the other hand, UAVs can also be used to breach public privacy and security, and sometimes UAV swarms may threaten sensitive facilities such as nuclear plants, oil installations, and large public gatherings. Thus, detecting, locating, and tracking UAVs is of great importance. A few popular techniques for UAV detection include the use of optical or acoustic radars~\cite{UAVdetection,UAVdetection1}, passive radio frequency~(RF) radars~\cite{UAVdetection2}, and MIMO radars~\cite{UAVdetection3}. These detection techniques have their distinct advantages/disadvantages and are chosen based on the need and type of a specific application, UAV size under consideration, among other criteria. For instance, small UAVs are harder to detect by radars as their radar cross-section (RCS) is similar to a bird~\cite{UAVbird} and their acoustic signature is typically weak to be detected by acoustic radars~\cite{UAVacoustic}. Moreover, a radar signal has to go through both the forward and reverse links, and hence it is subject to path loss in both directions, and suffers from reflection loss. Readily transmitted RF signals, on the other hand, when available, can be convenient signal sources to detect and track drones by using inexpensive passive RF sensors~\cite{UAVrf1, UAVrf2}. Such RF sensor network may have broader uses such as spectrum compliance monitoring and jammer identification, and unauthorized drone detection/tracking can be triggered whenever a drone signal in a tracking database is identified. It may also be possible to reuse an existing cellular network for RF sensing and drone detection purposes as studied in~\cite{Priyanka_paper}. We have used such an RF sensor in our study to detect and localize a flying UAV.

The majority of the UAVs use 2.4 GHz and 5.8 GHz ISM~(Industrial Scientific and Medical) frequency bands to communicate with the remote controllers. Therefore, an RF sensor can passively detect and localize the UAVs using the RF signals. In this work, we detect the RF signals and localize UAVs using the Keysight Geolocation system N6854A with the data collected by RF sensor N6841A. The sensor is capable of detecting RF sources in the vicinity of 1-2 km in sub-6~GHz bands, and inter-operates with a geolocation tool, N6854A~\cite{Geo}, which performs localization. Occasionally these RF sensors may perform poorly resulting in large localization errors as in~\cite{RF_sensor_defence}, where the authors utilized an N6841A sensor in a spectrum sensing module. These high errors are not acceptable for some critical applications. In this context, we propose an extended Kalman filter~(EKF) framework in this paper to improve tracking accuracy for GPS-denied environments.

EKF is a popular and effective approach for tracking UAVs in the literature. In \cite{mao2007design}, the authors propose an EKF based UAV localization technique using inter-UAV distance and the result shows that a GPS denied UAV can be localized with a maximum 40 m error. EKF can also be used for UAV collision avoidance as shown in the experimental study~\cite{luo2013uav}. The authors also present a comparison between EKF and least square (LS) estimation for UAV positioning. Both simulation and experimental study of indoor UAV localization are demonstrated in~\cite{li2018accurate} using two different EKF based algorithms. The simulation study achieves an error of lesser than 1 m for both algorithms. Similar to~\cite{li2018accurate}, an outdoor experiment is conducted in our work where EKF is used to improve the localization error.

The accuracy of the EKF algorithm relies on the dynamics of the UAV trajectory. One way of modeling the problem of localizing a highly maneuvering~UAV with an unknown pattern is to segmentize the trajectory and use multiple motion models (MMs) that represent the possible maneuvering patterns for each segment~\cite{liu2020interacting}. This approach assumes that the UAV follows one of the possible models and the precision of the model depends on the accuracy of this assumption. In this paper, we adopt a multiple MM-based EKF framework that considers three types of motion, i.e., moving in a circular trajectory and moving on a straight line with constant velocity or constant acceleration. More specifically, we label these three kinematic models as constant turn~(CT), constant velocity~(CV), and constant acceleration~(CA) models. A multi-MM framework models the trajectory better and, thus, provide better performance than a single-MM model.

The core purpose of this study is to analyze the performance of the EKF-based tracking with different MMs and its relationship to parameter selection (e.g., noise covariance) on a realistic dataset. The rest of the paper is organized as follows. In Section~\ref{Sec:Measurement}, we present the measurement setup, details of the Keysight geolocation system, the UAV trajectory, and the results after post-processing the data. The EKF implementation along with the modeling of multiple MM are presented in Section~\ref{Sec:localization}. The numerical results along with insights and discussions are given in Section~\ref{Sec:results}. Finally, we conclude the paper in~Section~\ref{Sec:conclusion}.

\section{Experiment Setup and Measurements}\label{Sec:Measurement}
In this section, we briefly describe the capabilities and data collected by the RF sensors along with its important parameter selection (e.g., noise covariance). This is followed by the presentation of UAV trajectory in an open field and the simulation results from the N6854A system for the post-processed data.

\subsection{RF Sensor System and the UAV}
In this work, commercially available Keysight RF sensor N6841A\footnote{These RF sensors are also being deployed as a part of the \textit{AERPAW: Aerial Experimentation and Research Platform for Advanced Wireless}~\cite{aerpaw}, \cite{aerpaw2}, a large-scale 5G testbed focused on wireless communications and UAVs.} is used for detecting, recording, and time-stamping the UAV downlink RF signals. The sensor node consists of RF sensor N6841A, a GPS antenna, and a broadband antenna as shown in Fig.~\ref{fig:RF_sensor}. The RF receiver operates in the frequency range of 20 MHz - 6 ~GHz and has a bandwidth up to~20 MHz. The broadband antenna is omnidirectional for better reception, and the GPS antenna helps to record the timestamps of the RF sensor collected data. This sensor system is also accompanied by a localization software called \textit{geolocation software N6854A}~\cite{Geo} that can localize RF sources within a~2~km radius. 

The geolocation software is capable of calculating time-frequency response, correlation amongst the RF sensors, and localization. For localization task, it offers three different algorithms, i.e., the time-difference-of-arrival (TDoA) based, the received-signal-strength (RSS) based, and a hybrid algorithm of the former two. The performance of these algorithms depends on the scattering environment. For instance, TDoA-based localization is preferred over RSS-based localization when the number of multipath components (MPCs) is low~\cite{tdoaRSS}. In a rich-scattering environment, the TDoA-based algorithm performs poorly, so the RSS-based algorithm is preferred. Our experiment was conducted in a large open field with a dominant line-of-sight (LOS) path and very weak reflections, hence, TDoA-based localization approach was preferred. Furthermore, we used a commercially available drone, a DJI Inspire 2, operating at 2.400 - 2.483 GHz band with a bandwidth of 20 MHz as the target that is controlled with an RF remote controller.

\subsection{RF Sensor Measurement Setup and UAV Trajectory}
Our experiment with Keysight RF sensors and the UAV was conducted in Dorothea Dix Park, Raleigh, North Carolina. The experiment field is shown in Fig.~\ref{fig:Google_Earth}, where white boxes are the RF sensors' location (black square points in Fig.~\ref{fig:UAV_Trajectory}). In total, four Keysight N6841A RF sensors were used to localize a UAV, and these sensors had a separation distance of at least 50 meters. For the system functionality, at least one of the RF sensors should be connected with a PC to monitor the UAV RF signals. These sensors can also be accessed via the internet when all of them are connected to the same LAN. Each of the RF sensors collects the RF signals' in-phase and quadrature (IQ) data. The collected data is post-processed by the geolocation software to estimate the UAV location using the TDoA-based algorithm. Upon performing localization, the geolocation software provides the latitude and longitude of the UAV in decimal degrees and does not provide the altitude information. Location estimates are logged  in conjunction with their respective timestamps.  

The UAV trajectory in the experiment based on the GPS data collected by the drone is shown in Fig.~\ref{fig:UAV_Trajectory} in 2-D. Apparent from the figure that the UAV was flown arbitrarily given the start and end point of the trajectory as pointed in the Fig.~\ref{fig:UAV_Trajectory}. We emphasize that such a trajectory cannot be modeled with a single MM (e.g., CV or CT) for the EKF framework to perform good. More details on the approximation of the trajectory are discussed in Section~\ref{Sec:localization}. Further, during the flight, the UAV records its own location and saves it (with the timestamps) in every 100~msec. This location information is used as the ground truth to evaluate the performance of the RF sensor system and the proposed EKF approach. Up next, we further explain data processing performed on the collected experimental data.

\begin{figure}[t!]{}
    \centerline{\includegraphics[scale=.8]{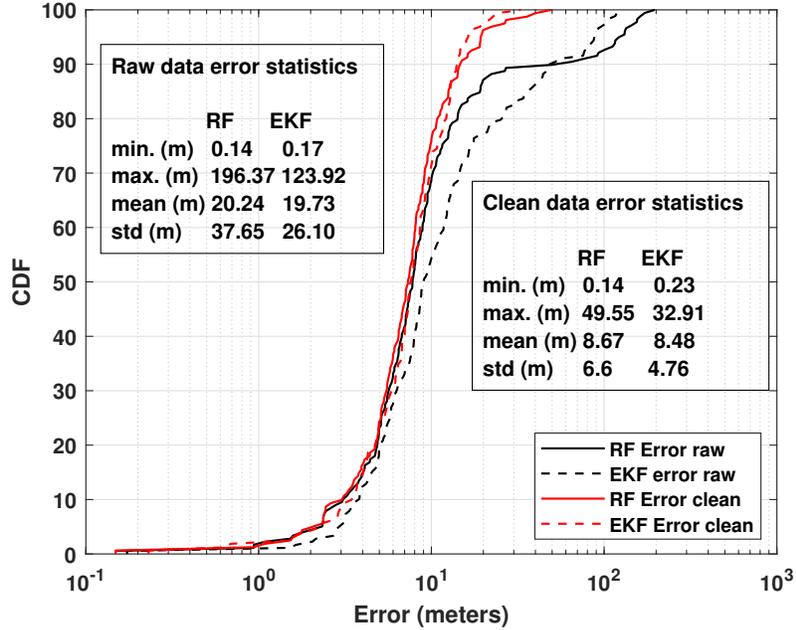}}
     \caption{Error CDF of the raw and clean data from the RF sensor and EKF against ground truth locations.}
     \label{fig:Ground_truth_cdf}
\end{figure}

\subsection{Data Post-processing and RF Sensor Error Performance}
With the experimental setup mentioned above, the UAV location estimate and the estimated locations from the RF sensor paired with corresponding timestamps were collected for around 5 minutes of flight time. The collected location information from both the UAV and RF sensors was converted from decimal degrees into cartesian coordinates for ease of representation. Further, the sampling time intervals of the UAV and RF sensor data are not equal\footnote{The number of data points from the RF sensors (sampling rate $\sim 1$~sample/sec) is lower than the number of data points at the UAV (sampling rate $\sim1$~sample/100msec).}. So to be able to work on a consistent dataset, we processed two datasets and filtered out samples that the timestamps do not match in millisecond sensitivity. After this process, we ended up with a total of $K = 178$ discrete timestamps. The processed UAV and RF sensor locations at $k^\text{th}$ time index are denoted by $\mathbf{l}^\text{UAV}_k$ and $\mathbf{l}_k^\text{RF}$, respectively. These data points are used for evaluating the performance of the geolocation~system.

The performance of the geolocation system is evaluated using the Euclidean distance metric. 
The black solid curve in Fig.~\ref{fig:Ground_truth_cdf} shows the cumulative distribution function (CDF) plot of the error between the RF sensor estimations and the UAV GPS data. We also report the error statistics (e.g., max, min, mean, and std) on the processed data points. Occasionally, the estimated location is far away from the true points corresponding to a higher error in the upper tail of the CDF (maximum of 198.4477 m). A possible explanation for this phenomenon is that the RF sensors would have intermittently detected other RF signal sources in a 1-2 km radius resulting in erroneous estimation. If the erroneous locations are used for further implementing EKF, the performance of the EKF will reduced as denoted by the dotted black line in Fig.~\ref{fig:Ground_truth_cdf}. It can be also observed from the same figure that the upper 10\% errors (more than 60 m) are abruptly higher than the lower 90\% errors. To disregard these erroneous data points, we cleaned the data by removing the points with an error higher than 60m. The resulting data points are referred as \textit{clean data}, whereas the data set with all the points are referred \textit{raw data}. The solid red curve in Fig.~\ref{fig:Ground_truth_cdf} shows the error CDF of the clean data. The highest error on this curve is around 50 m and the errors are greater than 14~m 90\% of the time. This much error might be too high depending on the application. The accuracy of the sensors can be improved by the use of an EKF framework, especially for the high error portions of the CDF plot. 

\section{Motion Model based EKF}\label{Sec:localization}
In this section, we first give details of the proposed EKF approach for a discrete-time system followed by the working principle of the MMs and segmentation of the trajectory.

\subsection{EKF Tracking Framework}
The primary objective of discrete EKF tracking is to estimate the present state given the information of the previous states in terms of time. Target tracking methods are usually model-based and are expressed in terms of the target dynamic MM and its observation model, respectively. Mathematically, a discrete-time model is expressed as follows~\cite{li2000survey}
\begin{eqnarray}\label{Eqn:equation1}
 \begin{aligned}
\textbf{State model:} \quad  & \mathbf{s}_{k}  = f_k ( \mathbf{s}_{k-1}, u_k) + \mathbf{w}_{k-1},\\
\textbf{Observation model:} \quad   & \mathbf{z}_{k}  = \mathbf{H}\mathbf{s}_{k} + \mathbf{v}_{k},
     \end{aligned}
 \end{eqnarray}
where, $k$ denotes the state (time) index $k \in \{1,\ldots,K \}$. In the state model, the term $\mathbf{s}$ denotes the state vector which consists of the parameters for predicting in each state. This representation of state is core to the EKF process and is basically the minimum information required about the past and present to determine the future response given the future input. The function $f_k$ is the vector-valued (time-varying) function that dictates the state transition. In the observation model, $\mathbf{z}_k$ and $\mathbf{H}$ denote the observations at the state index $k$ and observation matrix (not a time-varying function in our work as we obtain only position estimates from the RF sensor), respectively. Finally, the terms $\mathbf{w}_{k}$ and $\mathbf{v}_k$ are the process and observation noises that take into account any irrelevant attribute affecting the state and observations. Both $\mathbf{w}_{k}$ and $\mathbf{v}_k$ are assumed to be zero-mean Gaussian distribution with the covariances $\mathbf{Q}_k$ and $\mathbf{R}_k$ that control the level of relaxation of the constant-term MM assumptions and measurement noise level, respectively. Importantly, the choice of the state components $\mathbf{s}_k$, the function $f_k$ that accounts for the effect of target motion, and the covariance matrix $\mathbf{Q}_k$ depends on the choice of the motion-model~(MM).

\subsection{Kinematic Motion Models (MMs)}
To better model the random trajectory of the UAV, we used three tractable kinematic models, namely, constant velocity MM~(CV-MM), constant acceleration MM~(CA-MM), and constant turn MM~(CT-MM). The trajectory will be segmentized and only one of these MMs will be used for a single segment. Note that, unlike many other works in the literature, arbitrariness of our trajectory makes it impossible to model the whole trajectory with a single MM. The three considered MMs are described below. Readers can refer to~\cite{li2000survey} for further details.

\textbf{1) CV-MM:} The simplest form of motion is the CV-MM, where the speed and the direction of the object is constant, i.e., the object moves in a straight line. For this MM, the minimum information required in the state model to capture the future response are the current positions and the velocity. This is represented by~$\mathbf{s}_k$. The state vector and discrete-time linear state transition matrix, $f_k$ in~(\ref{Eqn:equation1}) are as follows \cite[Equation (16)]{li2000survey}
\begin{eqnarray*}
\begin{aligned}
    \mathbf{s}_k^\text{CV} & =  \begin{bmatrix} x \\ y  \\ \dot x \\ \dot y \end{bmatrix}, \quad \mathbf{f}_k^\text{CV}  =  \begin{bmatrix} 1 & 0 &  T_k  & 0 \\ 0 & 1 & 0  & T_k \\ 0 & 0 & 1  & 0  \\ 0 & 0 & 0  & 1  \end{bmatrix}, 
     \end{aligned}
\end{eqnarray*}
where the terms $\dot x$, $\dot y$, $T_k$ denote the velocity in $x$ and $y$ directions, and interval between the time-indices $k-1$ and $k$, respectively.  

\textbf{2) CA-MM:} The second considered MM is the CA-MM that has constant acceleration where the velocity is changing with a constant rate each second and is captured in the state vector by the terms $\ddot x$ and $\ddot y$ denoting the acceleration in $x$ and $y$ directions, respectively. The state vector and discrete-time linear state transition matrix, $f_k$ in~(\ref{Eqn:equation1}) are as follows~\cite[Equation (21)]{li2000survey}
\begin{eqnarray*}
\begin{aligned}
    \mathbf{s}_k^\text{CA} &=  \begin{bmatrix} x \\ y  \\ \dot x \\ \dot y \\ \ddot x \\ \ddot y \end{bmatrix}, \quad \mathbf{f}_k^\text{CA}  =  \begin{bmatrix} 1 & 0 & T_k  & 0 & \frac{T_k^2}{2}  & 0 \\ 0 & 1 & 0  & T_k  & 0 & \frac{T_k^2}{2}\\ 0 & 0 & 1  & 0 & T_k  & 0  \\ 0 & 0 & 0  & 1  & 0 & T_k \\ 0 & 0 & 0 & 0 & 1 & 0 \\ 0 & 0 & 0 & 0 & 0 & 1 \end{bmatrix}, 
     \end{aligned}
\end{eqnarray*}

\textbf{3) CT-MM:} This MM captures the object moving in a circular path with a constant angular velocity of $\omega$ (rad/sec). The state-space equations expressed in Cartesian~\cite{laneuville2013polar} are as follows:
 \begin{eqnarray*}
\begin{aligned}
    \mathbf{s}_k^\text{CT} & \! = \!  \begin{bmatrix} x \\ y  \\ \dot x \\ \dot y \\ \phi\end{bmatrix},  \mathbf{f}_k^\text{CT}(\mathbf{s}_k) \!  = \!  \begin{bmatrix} 
    x + \frac{\dot x}{\omega} \sin(\omega T_k) - \frac{\dot y}{\omega} (1 - \cos(\omega T_k))\\ \dot x \cos(\omega T_k) - \dot y \sin(\omega T_k)\\ 
    y + \frac{\dot x}{\omega}  (1 - \cos(\omega T_k))  + \frac{\dot y}{\omega}\sin(\omega T_k) \\ 
    \dot x \sin(\omega T_k) + \dot y \cos(\omega T_k)  \\ 
    \omega \end{bmatrix}.
\end{aligned}
\end{eqnarray*}

Regarding the noise factors, we assumed in this study that the process noise,$\mathbf{w}_{k}$, is uncorrelated across its components, and we adopted the covariance structure $\mathbf{Q}_k$ as explained in~\cite[Equation (17, 22)]{li2000survey}.

On the other hand, we only have the position information available from the RF sensor for the measurement noise, $\mathbf{v}_{k}$, and this is reflected by the measurement model for each of the MMs as follows
\begin{eqnarray*}
\begin{aligned}
    \mathbf{H}^\text{CV} \! & = \! \begin{bmatrix} 1 & 0 & 0 & 0 \\ 0 & 1 & 0  & 0  \end{bmatrix}, \quad    \mathbf{H}^\text{CA} \!  = \! \begin{bmatrix} 1 & 0 & 0 & 0 & 0 & 0 \\ 0 & 1 & 0  & 0 & 0 & 0 \end{bmatrix},\\
     \mathbf{H}^\text{CT} \! &= \!  \begin{bmatrix} 1 & 0 & 0 & 0 & 0 \\ 0 & 1 & 0  & 0 & 0  \end{bmatrix}.
     \end{aligned}
\end{eqnarray*}

Apart from the above-mentioned models, one might also consider other sophisticated MMs~\cite{mehrotra1997jerk} such as constant jerk, simultaneous deceleration, and turn, oscillatory turns, etc. However, we used only these three models as we found out that these three models are sufficient to model the whole trajectory. The crucial  part of the process is to decide how to do the segmentation of the trajectory and assigning an MM for each segment. In practice, this can be achieved using some data-driven estimators~\cite{linder2005non}, camera-based you-look-only-once~(YOLO) detection methods~\cite{ccintacs2020vision}, or leveraging environmental priors (physical constraints and environmental semantics) and motion history. However, for the sake of simplicity and tractability, we assumed that we perfectly know the segments and which MMs to be used apriori based on the GPS data.

\subsection{Implementation of EKF}
\begin{figure}[!t]
    \centering
    {\includegraphics[scale=.65]{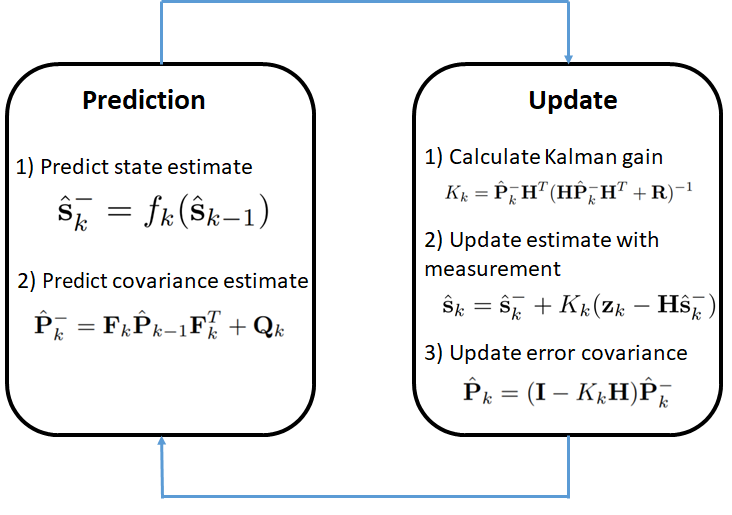}}
    \caption{EKF steps for each time index $k$.}
    \label{fig:ekfupdate}
\end{figure}

In this subsection, we discuss the details of the implementation of the EKF. The flow diagram is given in Fig.~\ref{fig:ekfupdate}. An EKF process is comprised of two consecutive steps, i.e., the prediction step and the update step. 

In the prediction step, the target state~$\hat{\mathbf{s}}_k$ and the covariance~$\hat{\mathbf{P}}^{-}_k$ at time index $k$ are predicted based on the kinematic process MMs, previous state $\hat{\mathbf{s}}_{k-1}$, and covariance estimate. These predictions will further be used in the update step, where the EKF updates the predicted state based on the Kalman gain $K_k$ and the measurement $\mathbf{z}_k$. The Kalman gain $K_k$ acts as a weighting factor that signifies the trustable value of the measurement and prediction state values. If the observations are assumed to be highly noisy (high variance), then $K_k$ will be low weighing less on the measurements and more on the predicted values, and vice-versa. Readers can refer to~\cite{Tu1} for further details. The above process is recursive and repeated for each time index $k$. All the implementation were carried out using the \textit{trackingEKF} built-in function of the \textit{Sensor Fusion and Tracking Toolbox} of Matlab\textregistered.

\begin{figure*}[!t]
    \begin{tabular}{cccccc}
    \includegraphics[scale=.28]{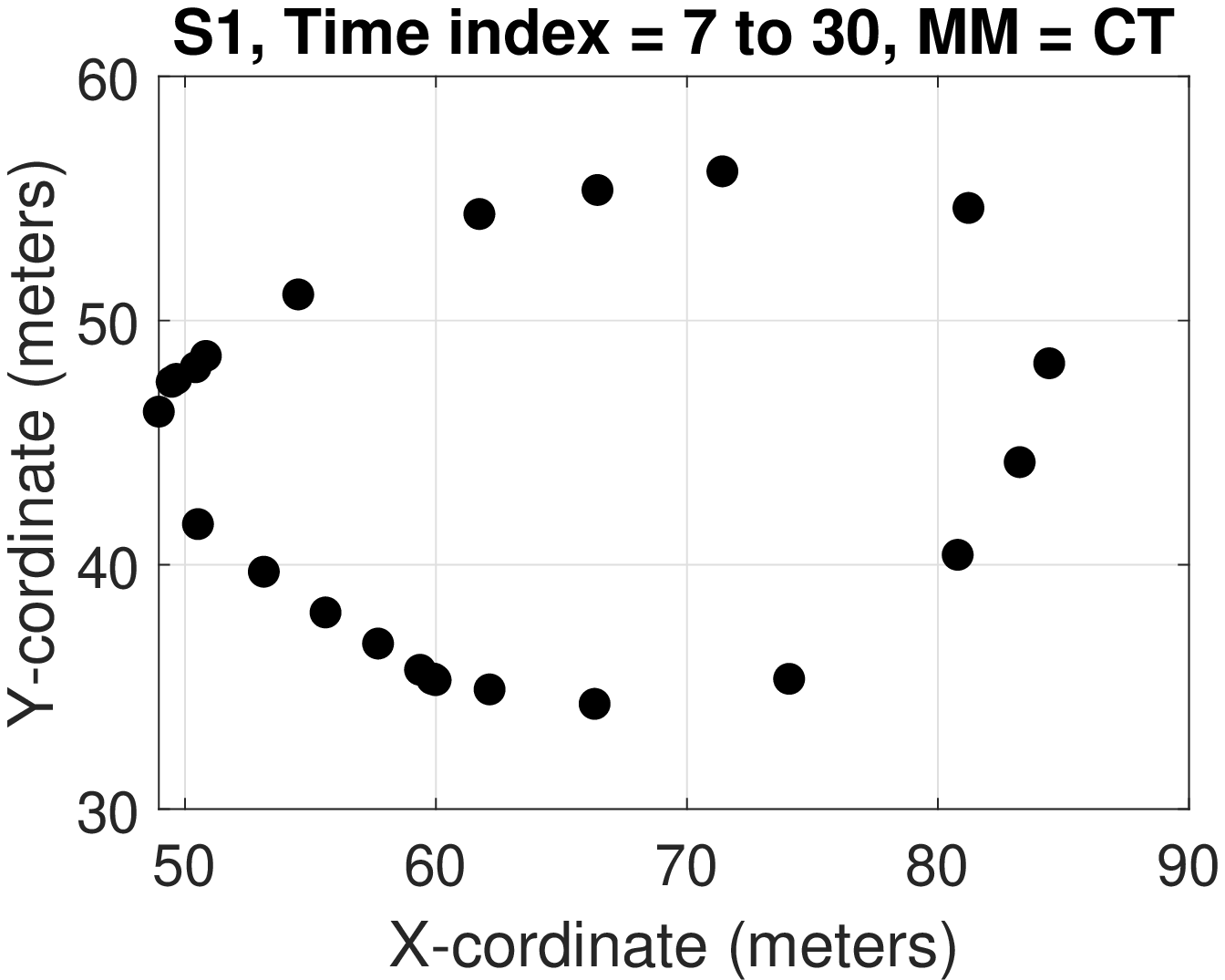}\!
    \includegraphics[scale=.28]{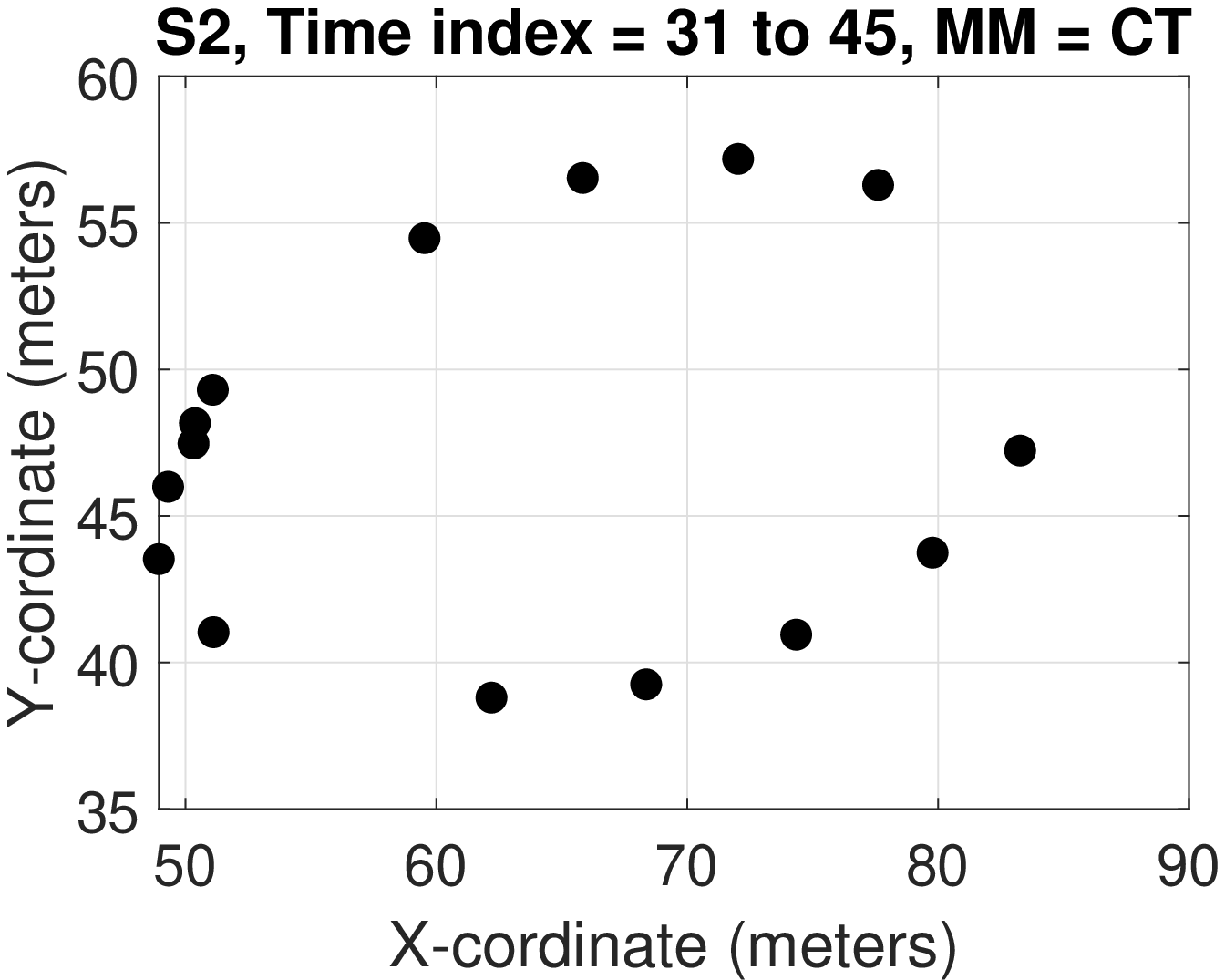}\!
    \includegraphics[scale=.28]{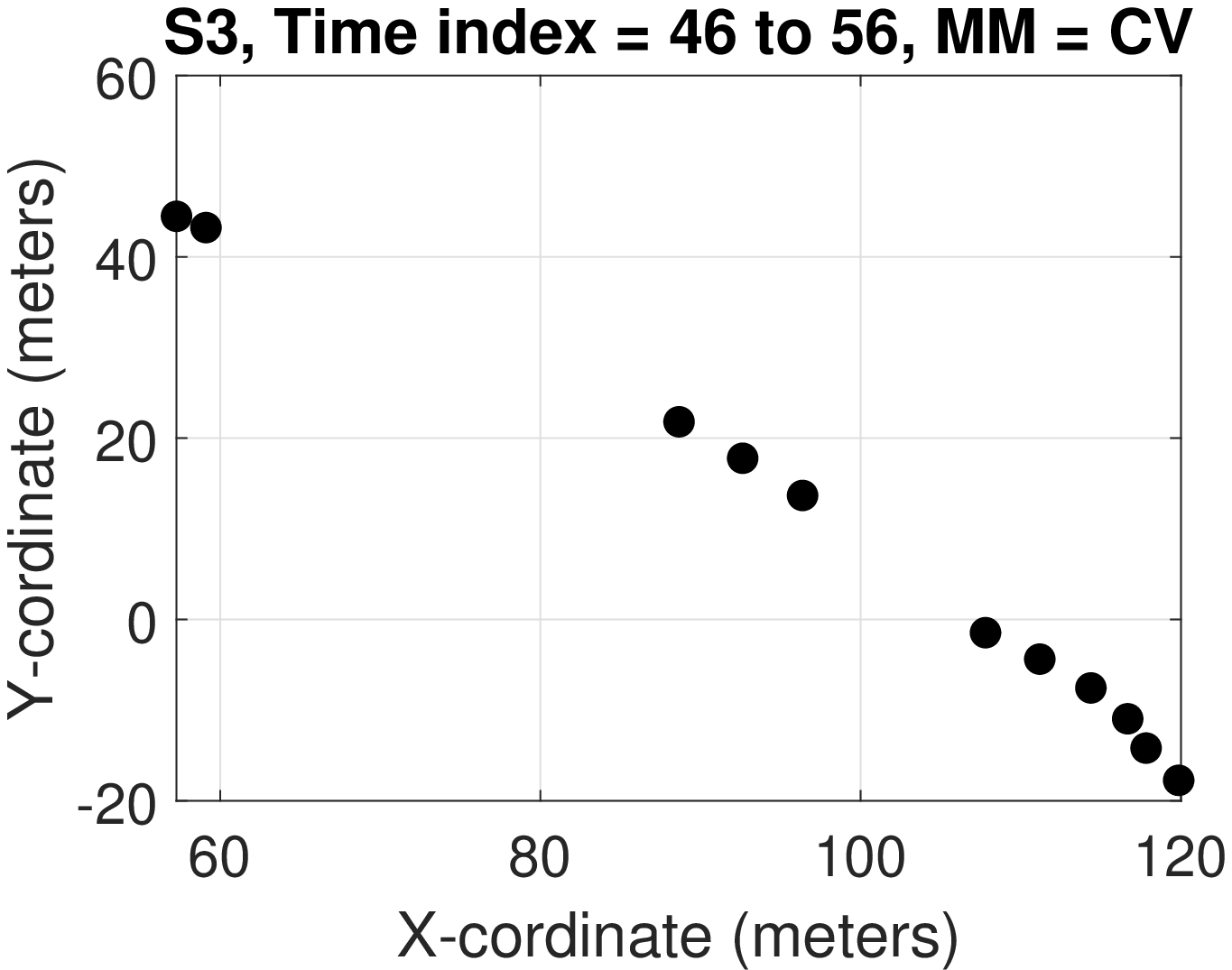}\!
    \includegraphics[scale=.28]{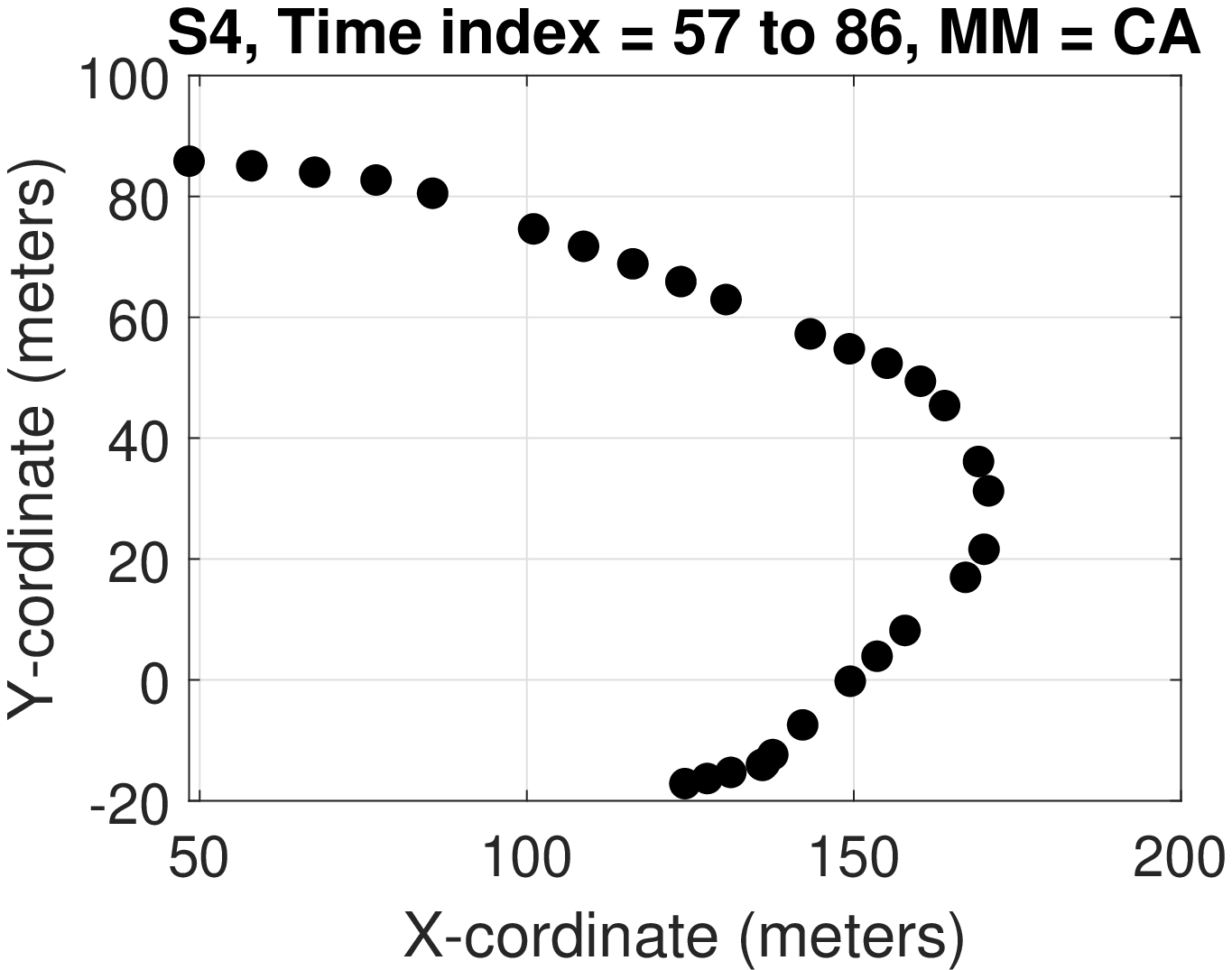}\\
    \includegraphics[scale=.28]{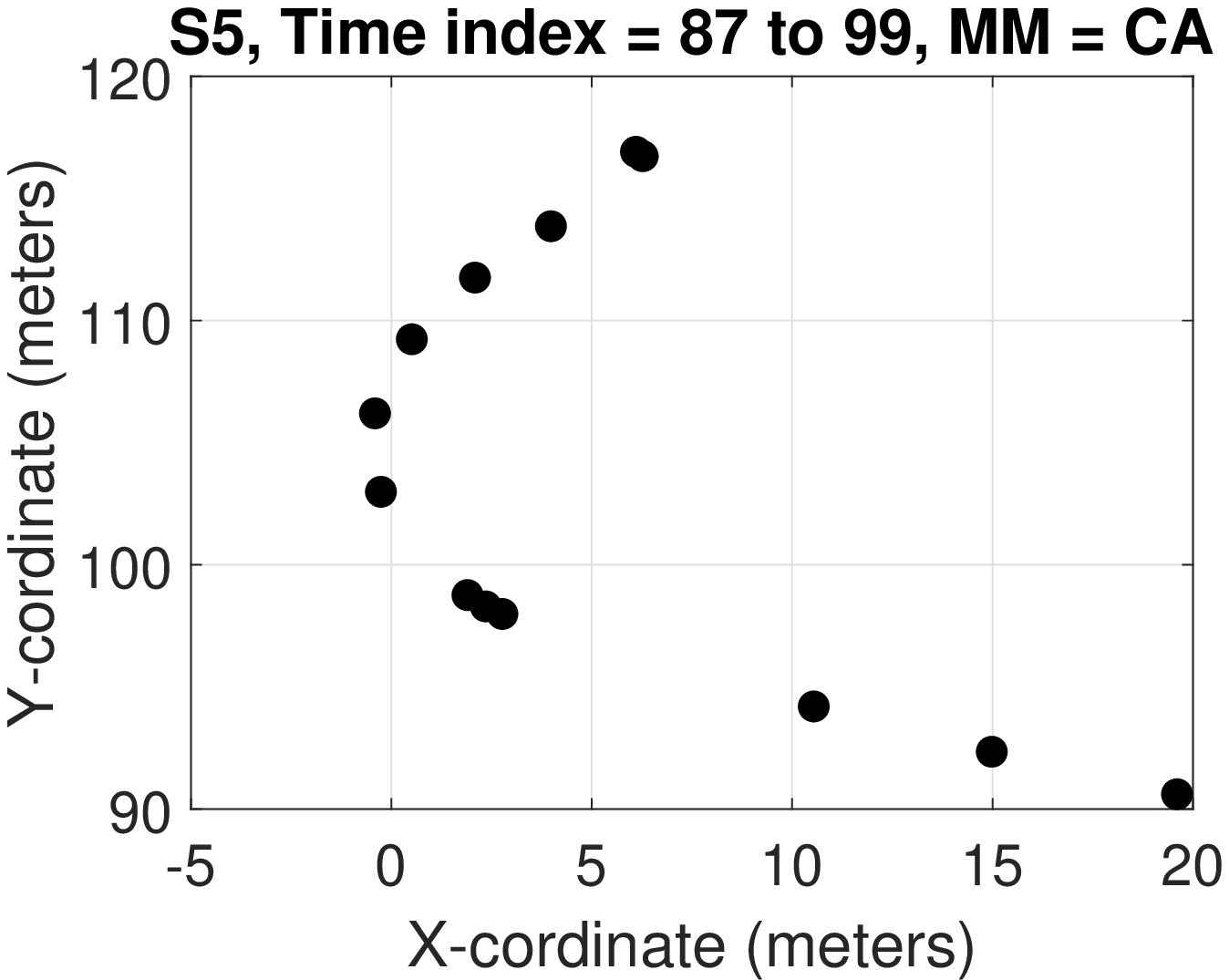}\!
    \includegraphics[scale=.28]{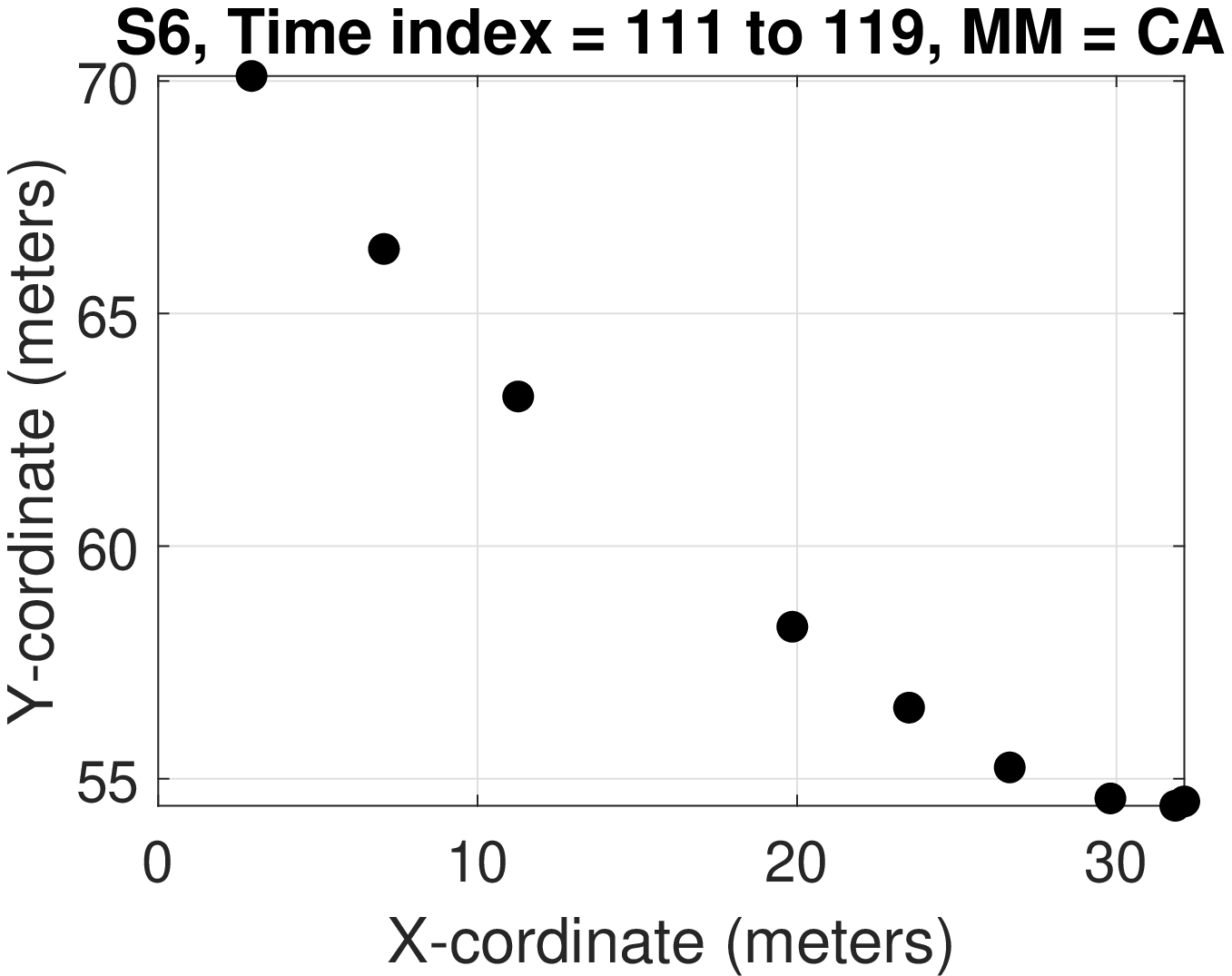}\!
    \includegraphics[scale=.28]{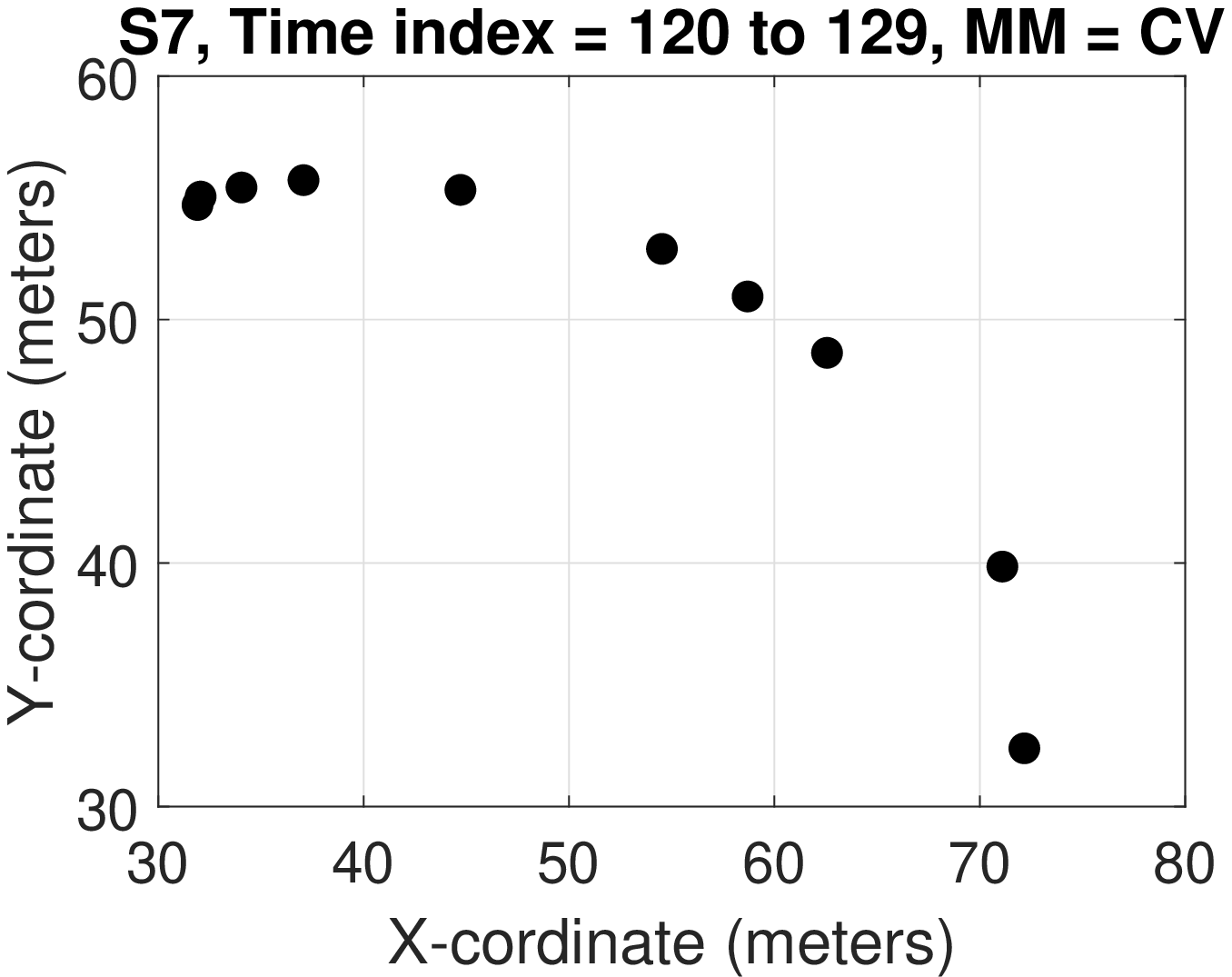}\!
    \includegraphics[scale=.28]{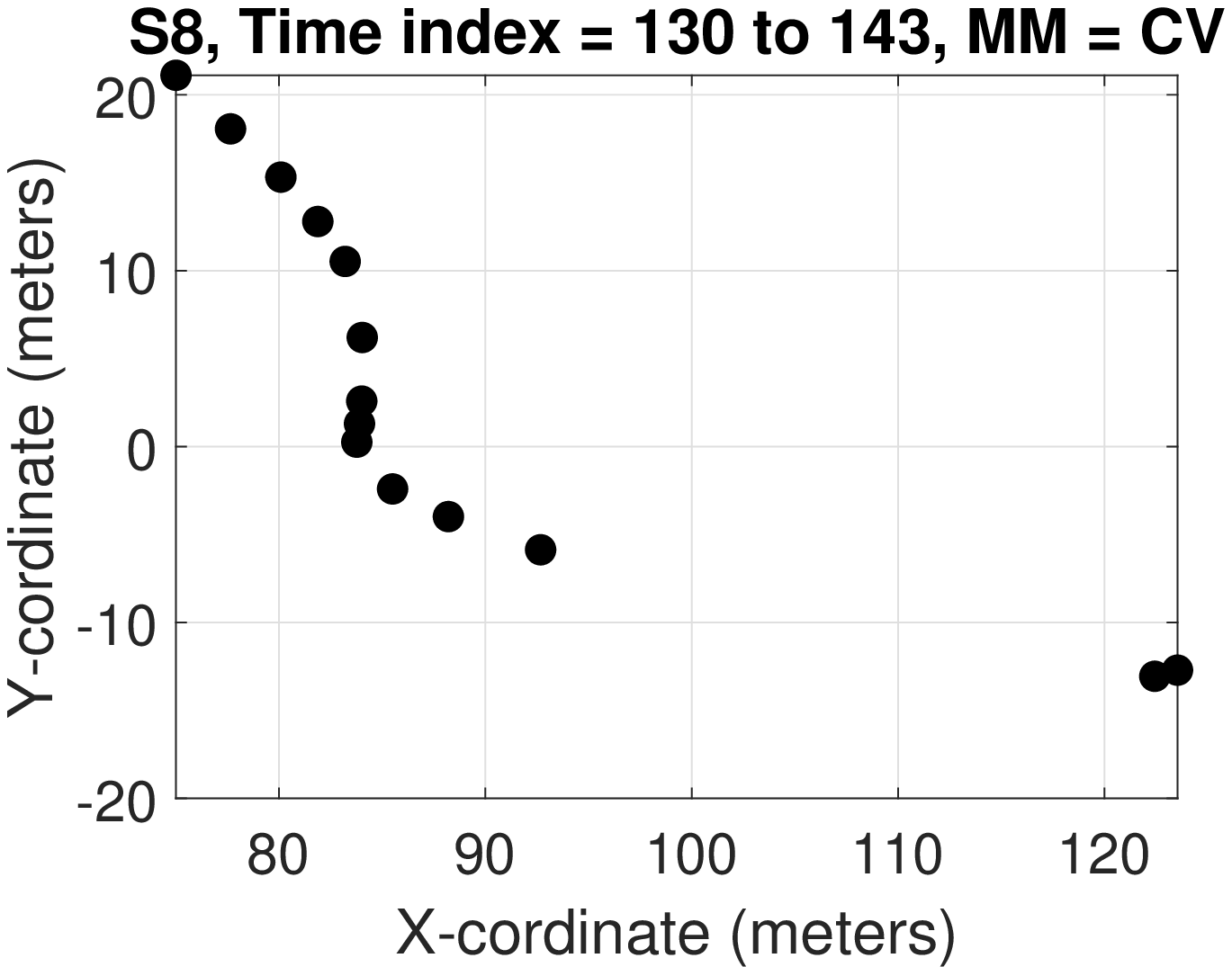}\\
    \includegraphics[scale=.28]{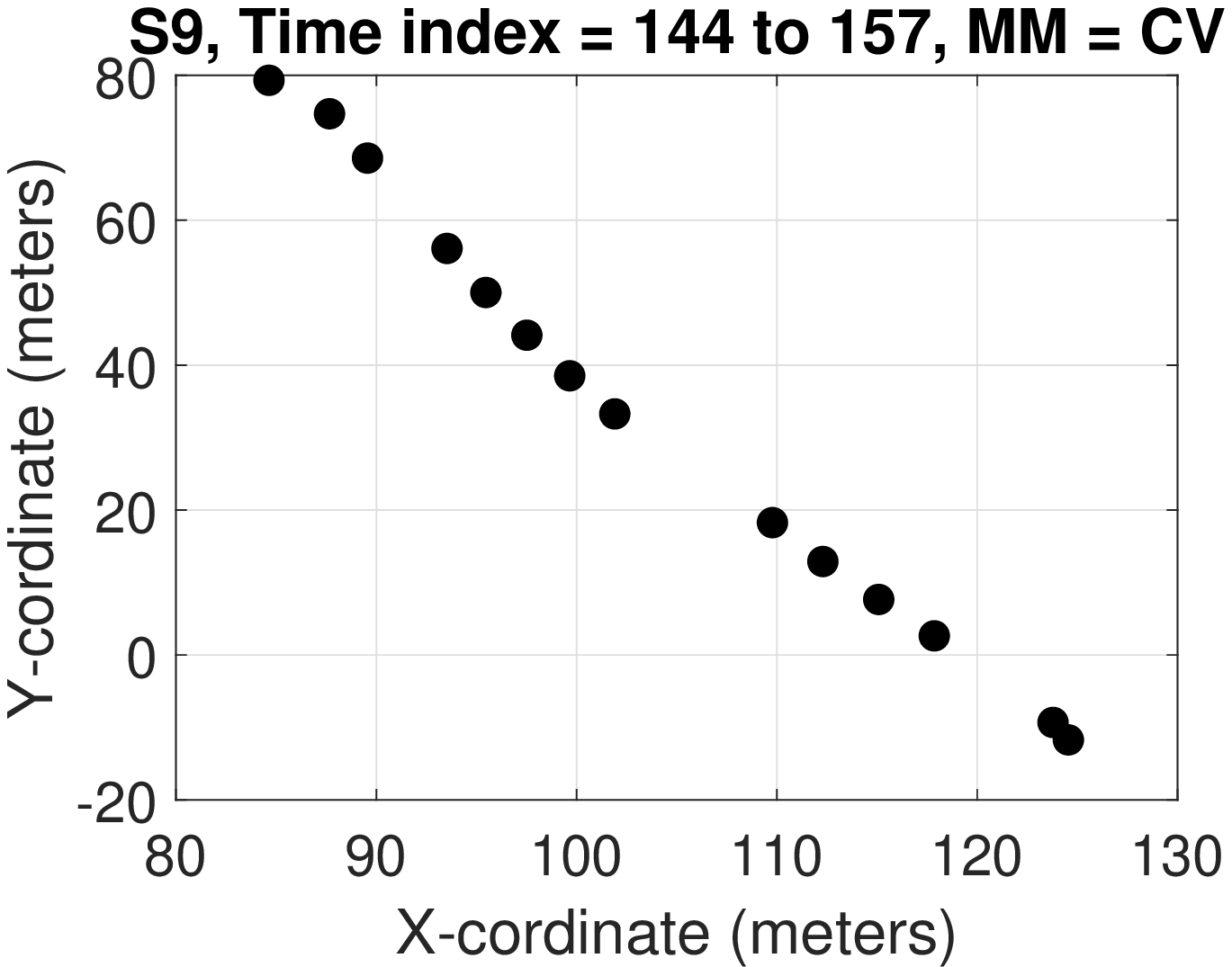}\!
    \includegraphics[scale=.28]{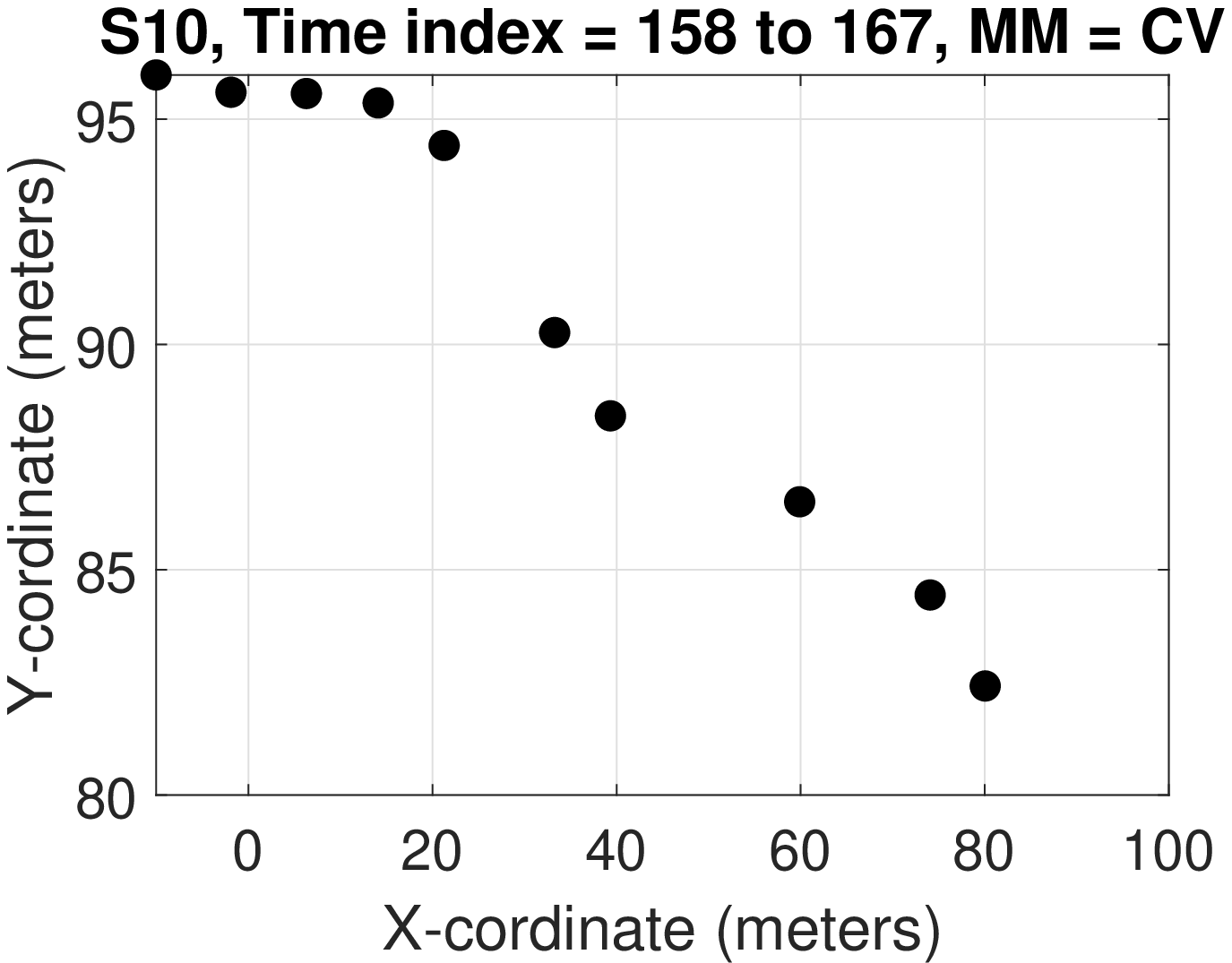}\!
    \includegraphics[scale=.28]{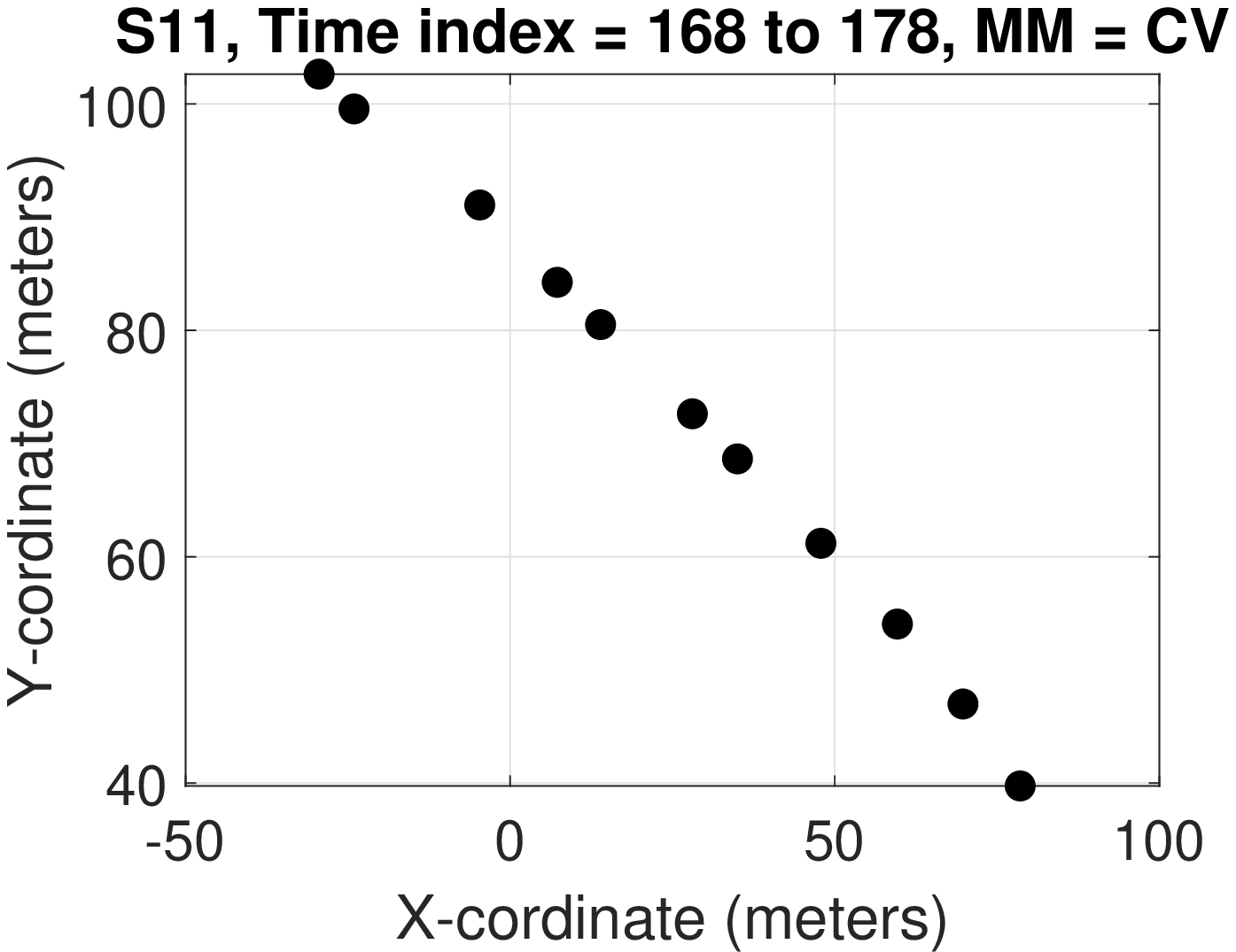}\!
    \end{tabular}
    \caption{Segmentation of the complete UAV trajectory into appropriate MMs.}
    \label{fig:segmentation}
\end{figure*}

\section{Experiment Results and Discussion}\label{Sec:results}
In this section, the performances of different MMs are observed on a realistic maneuvering UAV target. The statistics of the Euclidean distance metric $d_k(\mathbf{l}^\text{UAV}_k,\mathbf{l}^\text{Est.}_k)~=~|| \mathbf{l}^\text{UAV}_k - \hat{\mathbf{l}}^\text{Est.}_k||_2$ is used to evaluate the performance of both RF sensors and EKF. The terms $\hat{\mathbf{l}}^\text{UAV}_k$ and $\hat{\mathbf{l}}^\text{Est.}_k$ denote the ground truth (UAV location estimation) and estimated locations by RF and EKF~algorithms,~respectively.

\subsection{UAV Trajectory and Segmentation}
\begin{figure*}[!t]
         \centering
         {\includegraphics[trim= 1.75cm 0cm 0cm 0cm, clip,scale=.6]{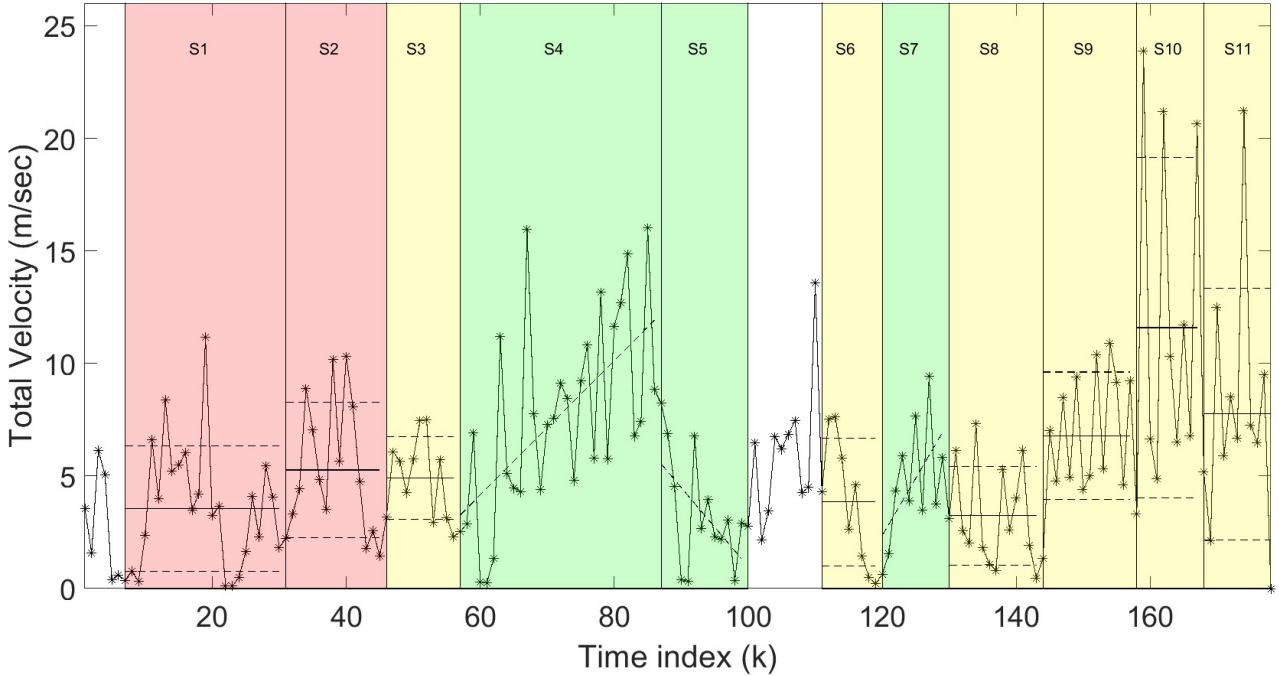}}
         \caption{Total velocity plot for clean UAV trajectory data where S1 - S2 : CT, S4, S5, S7 : CA, S3, S6, S8 - S11 : CV}
         \label{fig:VelocityPlot}
     \end{figure*}

The UAV trajectory flown for the experiment is as shown in Fig.~\ref{fig:UAV_Trajectory} with the red points being the measurement points and black line connecting the points with straight line. As noted in Section~II, after post-processing, we have a total of $K$ = 178 discrete time-stamps for the considered trajectory. First, we manually divide the complete trajectory into segments based on the visual inspection of the trajectory and velocity information obtained from the UAV, both are illustrated in Fig.~\ref{fig:segmentation} and Fig.~\ref{fig:VelocityPlot}. The trajectory knowledge and velocity variations within the segment would be sufficient to approximate a segment as either of the MMs~\cite{zhai2014constant}. As mentioned in Section 3.2, real-time trajectory segmentation is a different problem on its own. It requires either the trajectory or the environment knowledge apriori to segmentize the trajectory. For example, we can limit the velocity or acceleration variation if we have environmental knowledge. Further, test drives in the region of interest would provide an appropriate variance that can benefit EKF to perform well. In this work, we assume that an algorithm for the approximate segmentation of the trajectory is already available.
Fig.~\ref{fig:segmentation} shows the segmentation of the complete trajectory into eleven different stages of maneuvering along with its corresponding time indices and approximated MMs. Fig.~\ref{fig:VelocityPlot} shows the total velocity plot of the UAV trajectory for the complete trajectory according to the UAV estimated data. For ease of illustration, the CT, CA, and CV MMs are colored red, green, and yellow, respectively. These two figures complement each other and justify the segmentation. The initial UAV take off and the interim sections of the trajectory with many missing data points are colored white in Fig.~~\ref{fig:VelocityPlot} indicating that these segments are excluded. The number of data points in each segment (i.e., the length of each segment) varies across different segments, and it depends on how many points follow the MM definition for each segment.

Considering both Fig.~\ref{fig:segmentation} and Fig.~\ref{fig:VelocityPlot}, by intuition, first two segments (S1-S2) are approximated with CT-MM, and similarly, CA-MM is used for the segments S4, S5 and S7. The remaining segments, S3, S6, S8, S9, S10, and S11, are approximated as CV-MM due to approximated straight lines and velocities. Even though CV-MM is the best choice for this third segment group, EKF accuracy will reduce because of the deviation from idealized MM assumption due to high variations in velocity. The mean and standard deviation of velocity for CV and CT segments and those of acceleration for CA segments are also shown in Fig.~\ref{fig:VelocityPlot}. 

     
\begin{figure*}
\begin{center}
\captionof{table}{Error statistics for each segment (Clean data, Error $<$ 60 m).}
\label{tab:tab3}
\begin{tabular}{|c|c|c|c|c|c|c|c|c|c|c|c|c|}
\hline
\cellcolor[HTML]{FFFFFF}\textbf{} & \multicolumn{2}{c|}{\cellcolor[HTML]{FD6864}\textbf{S1}} & \multicolumn{2}{c|}{\cellcolor[HTML]{FD6864}\textbf{S2}} & \multicolumn{2}{c|}{\cellcolor[HTML]{FFFE65}\textbf{S3}} & \multicolumn{2}{c|}{\cellcolor[HTML]{67FD9A}\textbf{S4}} & \multicolumn{2}{c|}{\cellcolor[HTML]{67FD9A}\textbf{S5}} & \multicolumn{2}{c|}{\cellcolor[HTML]{FFFE65}\textbf{S6}} \\ \hline
\textbf{Err. stats.}              & \textbf{RF}               & \textbf{EKF}                 & \textbf{RF}                & \textbf{EKF}                & \textbf{RF}                & \textbf{EKF}                & \textbf{RF}               & \textbf{EKF}                 & \textbf{RF}               & \textbf{EKF}                 & \textbf{RF}                & \textbf{EKF}                \\ \hline
\textbf{min. (m)}                 & 1.08                      & \textbf{0.23}                & 0.93                       & \textbf{0.35}               & 5.3                        & 5.65                        & 0.14                      & 4.2                          & 5.54                      & 7.4                          & 1.53                       & \textbf{0.68}               \\ \hline
\textbf{max. (m)}                 & 14.25                     & \textbf{13.88}               & 9.04                       & 9.73                        & 15                         & 14.84                       & 39.89                     & \textbf{19.89}               & 49.55                     & \textbf{32.91}               & 7.32                       & \textbf{7.3}                \\ \hline
\textbf{mean (m)}                 & 7.7                       & 8.0                          & 5.89                       & 6.39                        & 8.89                       & 8.97                        & 10.59                     & \textbf{9.67}                & 14.63                     & \textbf{13.84}               & 4.72                       & \textbf{4.7}                \\ \hline
\textbf{std (m)}                  & 3.64                      & \textbf{3.49}                & 2.28                       & 2.59                        & 3.23                       & \textbf{2.79}               & 7.7                       & \textbf{3.8}                 & 11.87                     & \textbf{6.89}                & 2.13                       & 2.49                        \\ \hline
\end{tabular}
\end{center}
\end{figure*}

\begin{figure*}
\begin{center}
\captionof{table}{Error statistics for each segment (Clean data, Error $<$ 60 m).}
\label{tab:tab4}
\begin{tabular}{|c|c|c|c|c|c|c|c|c|c|c|}
\hline
\cellcolor[HTML]{FFFFFF}{\color[HTML]{333333} \textbf{}} & \multicolumn{2}{c|}{\cellcolor[HTML]{67FD9A}\textbf{S7}} & \multicolumn{2}{c|}{\cellcolor[HTML]{FFFE65}\textbf{S8}} & \multicolumn{2}{c|}{\cellcolor[HTML]{FFFE65}\textbf{S9}} & \multicolumn{2}{c|}{\cellcolor[HTML]{FFFE65}\textbf{S10}} & \multicolumn{2}{c|}{\cellcolor[HTML]{FFFE65}\textbf{S11}} \\ \hline
\textbf{Err. stats.}                                     & \textbf{RF}               & \textbf{EKF}                 & \textbf{RF}                & \textbf{EKF}                & \textbf{RF}               & \textbf{EKF}                 & \textbf{RF}                & \textbf{EKF}                 & \textbf{RF}                & \textbf{EKF}                 \\ \hline
\textbf{min. (m)}                                        & 2.33                      & \textbf{1.69}                & 2.39                       & \textbf{1.41}               & 6.46                      & \textbf{3.17}                & 1.75                       & 2.28                         & 2.04                       & \textbf{2.02}                \\ \hline
\textbf{max. (m)}                                        & 35.81                     & \textbf{23.86}               & 14.67                      & 14.67                       & 27                        & \textbf{19.53}               & 17.06                      & \textbf{12.84}               & 16.17                      & 16.31                        \\ \hline
\textbf{mean (m)}                                        & 7.68                      & \textbf{5.81}                & 5.99                       & 6.54                        & 10.72                     & \textbf{10.3}                & 8.23                       & \textbf{7.02}                & 7.36                       & 7.52                         \\ \hline
\textbf{std (m)}                                         & 9.94                      & \textbf{6.46}                & 3.97                       & 4.13                        & 5.66                      & \textbf{4.24}                & 4.96                       & \textbf{3.37}                & 5.1                        & 5.15                         \\ \hline
\end{tabular}
\end{center}
\end{figure*}
\subsection{EKF Parameters Selection Strategy}
While building the EKF model for each segment, the initial point was chosen to be the location measured by the RF sensor system. These estimates are noisy and the noise level is characterized by a covariance matrix $\mathbf{R}_k = diag\{\sigma^2_{x},\sigma^2_{y}\}$, where $\sigma^2_{x}$ and $\sigma^2_{y}$ values are chosen as the mean of the RF sensor location error. On the other hand, the process noise covariance matrix elements depend on the MM and are treated as design parameters. These elements are the velocity variance~$\sigma^2_{\dot x}$ and~$\sigma^2_{\dot y}$, acceleration variance~$\sigma^2_{\ddot x}$ and~$\sigma^2_{\ddot y}$, and variance of turn-rate $\sigma_{\omega}^2$. We keep these parameters fixed for each segment using the variance of the respective domain segments obtained from the UAV data. Ideally, these parameter values should be predefined from the system model definition. In this work, considering the lack of system model knowledge, the current values are chosen to be sufficient for our purposes in this experimental study.
However, the fixed values may not work under dynamic input conditions where the measurements (RF sensor output) vary due to external factors. Under such conditions, learning modules can be used to change the covariance at each step~\cite{ullah2019improving} for optimized performance. We will use this approach in our future work.

\subsection{Performance Analysis}
We applied EKF separately to each segment and studied the error performance of the RF sensor and the proposed tracking EKF algorithms. The error CDF curves for both \textit{raw data} and \textit{clean data} are also given in Fig.~\ref{fig:Ground_truth_cdf}. The black and red solid lines represent the error of the \textit{raw data} and \textit{clean data}, whereas, the dashed black and red lines are the error CDFs of the EKF predicted data. As expected, the results are better for the \textit{clean data} since EKF can reduce the highest error without disrupting errors of the rest much. On the other hand, the highest error is reduced significantly for the \textit{raw data} with the cost of an increase in errors of the remaining points. This is because the noise varies highly from point to point for some segments, and it is hard for EKF to perform well for all the points with fixed noise covariance. For the error analysis in each segment, we focus only on the \textit{clean data} and compare the maximum, minimum, mean, and standard deviation~(std) of calculated errors, and highlight the lowest error. These values are listed Table~\ref{tab:tab3}, and~\ref{tab:tab4}.

For both CT-MM segments (S1 and S2), the maximum error with the EKF is reduced significantly compared to the RF sensor even if it does not take into account the dynamic modeling. Thus, even an approximated MM-based EKF can help to reduce the highest error significantly. The EKF even helps reduce the minimum and standard deviation error for S1. For S2, the results are not improved but rather increased. This might be due to the nonalignment of the points with the circular assumption and velocity variation. Note that CT-MM assumes a constant angular velocity. A similar trend can be noticed for CA-MM and CV-MM segments. Among all the CA-MM, S7 performs the best with the reduction in all the observed error statistics. In S4 and S5, all other error statistics are reduced except the minimum error. Comparing all three CA-MMs, the acceleration variation is the smallest in S7 as it is closer to a straight line, hence, performing the best. In CV-MM case, S9 is mostly satisfying the CV-MM assumptions and all the error statistics are reduced. The segments S6 and S10 performs the second best because of the good alignment with the straight-line assumption. In S11, on the other hand, the points are well aligned on a straight line but the velocity variance is quite high resulting in no improvement by the EKF framework. Finally, segments S3 and S8 are performing the worst by uplifting the error statistics due to the discontinuity in data points and misalignment from the straight line.

To sum up, even though the proposed EKF approach does not give the lowest errors in all circumstances, it reduces the highest error significantly for different types of motions model as long as the segment trajectory is in compliance with the MM assumption. We also observe that the performance of the EKF depends on the noise covariance selection. Different noise covariances may improve errors at some particular points deteriorating at other points. One possible approach for this is to tune the EKF parameters at each state according to some prior knowledge of the change of motion or the environmental knowledge. A reinforcement learning-based approach can be adopted as mentioned in~\cite{ullah2019improving} by using an artificial neural network to update the EKF parameters according to the output and external factors.

\section{Conclusion}\label{Sec:conclusion}
In this paper, we proposed an EKF-based approach to improve the localization performance of Keysight N6854A (RF) sensors. The proposed technique has been verified on a dataset collected from a real-world experiment. As the chosen trajectory was arbitrary, it is sliced into multiple segments and EKF is applied separately to each segment with the best fitting motion model. Although perfect segmentation is not possible for an unplanned trajectory, the results of this study show that EKF can improve the error statistics when the points are well aligned with the assumed MM, and error can increase if opposite. 
We also observe that the performance of the model depends on chosen EKF parameters, i.e. noise covariance. Assuming that prior knowledge of the velocity distribution of each segment is in possession, the error can be reduced by varying the noise covariance at each state. A possible future direction for this work is to implement a machine-learning algorithm to choose the noise covariance dynamically. Another possible future work can be glitch detection and elimination of the RF sensor data provided a few measurement points are from nearby other RF sources. Overall, the UAV trajectory segmentation and fitting multiple motion model EKF can improve the localization performance of Keysight N6854A RF sensors provided the knowledge of approximate segments along with the respective motion models are known.

\section{Acknowledgment}
This work was in part supported by the NSF PAWR program under grant number CNS-1939334 and by the INL Laboratory Directed Research Development (LDRD) Program under DOE Idaho Operations Office Contract DEAC07-05ID14517.

\bibliographystyle{IEEEtran}
\bibliography{IEEEabrv,ref}

\end{document}